\documentclass[aps,prx,twocolumn,superscriptaddress,showpacs,floatfix]{revtex4-1}
\usepackage{amsmath,amssymb,amsfonts,amsthm}
\usepackage{braket}
\usepackage{graphicx}
\usepackage{bm}
\usepackage{bbm}
\usepackage{color}
\usepackage{dcolumn}   
\usepackage{epstopdf}
\usepackage[caption=false]{subfig}
\usepackage{xr}
\usepackage{color}
\usepackage{bbold}
\usepackage[colorlinks=true,linkcolor=blue,urlcolor=blue,citecolor=blue]{hyperref}
\usepackage{comment}
\usepackage{ulem}

\begin{document}
 
 
\title{Majorana zero modes in superconductor-magnet heterostructures with d-wave order}
\author{Bastien Fajardo }
\affiliation{Physics Department, McGill University, Montr\'eal, QC, Canada H3A 2T8}
\author{T. Pereg-Barnea}
\affiliation{Physics Department, McGill University, Montr\'eal, QC, Canada H3A 2T8}
\author{Arun Paramekanti}
\affiliation{Department of Physics, University of Toronto, 60 St.~George Street, Toronto, ON, Canada M5S 1A7}
\author{Kartiek Agarwal}
\affiliation{Material Science Division, Argonne National Laboratory, Lemont, IL, USA 60548}
\begin{abstract}
Magnetic skyrmions in proximity to superconductors offer a route to engineering topological superconductivity due to the synthetic spin–orbit coupling 
engendered by the spin twist of the skyrmion texture. Previous theoretical works show that this leads to Majorana zero modes (MZMs) 
in skyrmion-vortex pairs 
for $s$-wave superconductors. Here we investigate this mechanism in fully gapped $d+is$ and $d+id$ superconductors. We find the surprising result
that while stable MZMs are found in large parts of the phase diagram, strongly enhanced $d$-wave pairing or stronger skyrmion-induced spin twisting can in fact destroy topology unlike in $s$-wave superconductors. This effect can be understood from the non-trivial spatial structure of the $d$-wave pairing, and mixing of odd and even angular-momentum pairing channels in a rotated frame which untwists the skyrmion texture. Our results inform the feasibility of realizing MZMs with unconventional superconductors in such heterostructures.
\end{abstract}

\maketitle
\section{Introduction}

Topological superconductors hosting Majorana zeros modes (MZMs) remain central to ongoing efforts towards realizing quantum computers that are fault-tolerant at the hardware level~\cite{aasen2025roadmap,microsoft2025interferometric,Alicea2012,SatoAndo2017}. Among the various proposed platforms, hybrid structures combining conventional superconductors with engineered one dimensional (1D) magnetic chains or 2D magnetic textures have emerged as 
versatile systems in which effective spin-orbit coupling and topological superconductivity can be induced in otherwise trivial materials~\cite{martin2012majorana,klinovaja2013giant,Yang2016,mascot2021topological}. 2D magnetic skyrmions---nanoscale, topologically nontrivial spin textures---are especially attractive in this context due to their stability, tunability, and rich dynamical properties~\cite{NagaosaTokura2013,Wiesendanger2016}. 

In pioneering theoretical work, Yang et al.~\cite{Yang2016} showed that a magnetic skyrmion exchange-coupled to an $s$-wave superconductor can imprint a spatially varying spin--orbit field on the electrons, leading to a topological superconducting phase that supports a Majorana zero mode in the skyrmion core, provided that the skyrmion has an {\it even} azimuthal winding number. 
%
%
Experimentally, however, skyrmions observed in chiral magnets and multilayer heterostructures almost universally carry odd topological charge, most commonly $n_{\mathrm{sk}} = 1$~\cite{NagaosaTokura2013,Wiesendanger2016} though theoretical models stabilizing higher winding numbers are known~\cite{leonov2015multiply,ozawa2017zero}. 
This apparent incompatibility can be resolved by considering composite topological defects formed by binding a conventional $hc/2e$ 
superconducting vortex to a skyrmion of winding number one~\cite{Rex2019}. 
In such a skyrmion--vortex pair, Berry phases originating from the superconducting and magnetic vorticities add up, allowing MZMs even when $n_{\mathrm{sk}} = 1$. This scenario is not only theoretically appealing, but also experimentally relevant in light of recent demonstration of skyrmion--(anti)vortex coupling and controlled skyrmion textures in magnet--superconductor heterostructures~\cite{Petrovic2021,xie2024visualization}.

The need for superconducting vortices naturally directs attention to type-II superconductors; layered cuprate superconductors are especially appealing, since they offer large superconducting gaps, potentially enabling MZMs that are robust at significantly elevated temperatures. In certain cases, they show $T_c$ comparable to bulk $T_c$ even in the monolayer limit~\cite{yu2019high,logvenov2009high}, potentially allowing for stronger exchange couplings. However, nodal $d_{x^2-y^2}$-wave superconductors are intrinsically gapless, and low-energy quasiparticles compromise topological protection~\cite{vakili2023majorana}. 
%
Fully gapped unconventional superconducting states of the form $d+is$ or chiral $d+id$ can circumvent this issue--- 
such multi-component order parameters have been discussed extensively as candidates in various unconventional superconductors, including cuprates~\cite{SigristUeda1991} and other materials that exhibit time-reversal-symmetry-breaking superconductivity~\cite{Gong2017,Kapitulnik2009review}. 
%
Recently, twisted cuprate bilayers have emerged as a particularly promising platform to realize fully gapped, time-reversal-symmetry-breaking superconductivity at elevated temperatures~\cite{can2021high}. Experiments on twisted Bi$_2$Sr$_2$CaCu$_2$O$_{8+\delta}$ (Bi-2212) junctions~\cite{ZhaoTwisted2021} report  
direct signatures of time-reversal-symmetry-breaking~\cite{ZhaoTRSB2023}. 
These developments motivate us to explore the interplay between skyrmion textures and vortices of $d+is$ or $d+id$ superconductors in 
layered heterostructures.

In this work we investigate, using exact diagonalization of Bogoliubov--de Gennes tight-binding models, how MZMs are realized in superconducting--magnetic heterostructures that combine skyrmion textures with vortices in fully gapped $d+is$ and $d+id$ superconductors. Our analysis reveals a striking departure from the familiar $s$-wave case: increasing either the strength of the $d$-wave component or the effective spin--orbit coupling generated by the skyrmion (which depends on rate of winding of the skyrmion spin texture in the radial direction) can drive the system into a topologically trivial phase.  This behavior contrasts sharply with standard $s$-wave setups, where larger spin--orbit coupling and superconducting gaps are typically beneficial for stabilizing Majorana zero modes at vortices or wire ends~\cite{Alicea2012,SatoAndo2017,Yang2016}.

We trace the origin of this finding to the nontrivial transformation properties of the $d$-wave order parameter in the presence of a spatially winding exchange field. By performing a local spin rotation that aligns the skyrmion exchange field along a uniform $z$ axis, a spatially uniform spin--orbit coupling is generated, as in the $s$-wave case~\cite{Yang2016}, but the superconducting gap acquires a highly nonlocal structure in real and momentum space. In particular, the transformed $d$-wave pairing decomposes into a superposition of pairing channels with both even and odd orbital angular momentum. As a consequence, the usual topological criterion---namely, that the total winding coming from the superconducting vortex, the skyrmion texture, and the superconducting order parameter must be even to support a localized Majorana mode~\cite{Yang2016,Rex2019}---is not directly applicable. When the $d$-wave pairing amplitude or skyrmion-induced spin--orbit coupling becomes too large, the odd-angular-momentum components dominate, and the system is driven into a trivial superconducting phase despite the presence of vortices and skyrmion winding.

Our results reveal a subtle but essential constraint on engineering MZMs in magnetically textured superconductors with unconventional pairing: in contrast to the $s$-wave case, stronger $d$-wave components and skyrmion-induced spin twisting do not necessarily favor topology. Instead, a delicate balance between the amplitudes and relative phases of the multicomponent order parameter and the magnetic texture is required to maintain an effectively even total winding and a robust Majorana bound state. These insights inform the design of future experiments in cuprate-based and other layered superconducting heterostructures that aim to harness skyrmions and vortices for topological quantum devices.

The remainder of this paper is organized as follows. In Sec.~II we introduce the microscopic model for the superconducting--magnetic heterostructure, specifying the skyrmion texture, vortex configuration, and superconducting order parameters studied, and we outline the exact-diagonalization approach used to obtain the quasiparticle spectrum. In Sec.~III we present the numerical results for unconventional superconductors, including phase diagrams, and the evolution of MZMs as a function of the rate of radial winding of the skyrmion spin texture, the strength of exchange coupling, and $d$-wave amplitude. In Sec.~IV we analyze the problem in the rotated-spin basis where the skyrmion exchange field is uniform, elucidating how the transformed $d$-wave pairing generates mixed angular-momentum channels and modifies the topological winding condition. In Sec.~V we summarize the main conclusions, discuss their implications for ongoing experimental efforts in skyrmion--superconductor heterostructures and twisted cuprate systems, and suggest directions for future work.

\section{Model}\label{sec:Model}

\subsection{Skyrmion texture}

A magnetic skyrmion is a topologically nontrivial spin configuration in which the magnetization at the core points, for definiteness, along the $+z$ direction, while far from the core it aligns with $-z$. Thus, the $z$-component of the spin interpolates smoothly from positive to negative as a function of the radial coordinate. When this interpolation occurs over a short distance, the core resembles a simple domain wall. However, in a realistic skyrmion, the transition region acquires a finite thickness in which the spins lie predominantly in the plane and wind around the core. In this work we focus on N\'eel-type skyrmions, where the in-plane component of the magnetization points radially. 


The strength of the effective spin-orbit coupling realized via an exchange coupling to such a skyrmion depends inversely on the radial length scale at which the spin flips. For a skyrmion with a single flip of the spin from $+z$ to $-z$, while a shorter length scale can be expected to result in a larger spin-orbit coupling and concomitantly more tightly localized MZMs, this is counteracted by the smaller size of the skyrmion itself, which leads to stronger coupling between the pair of MZMs realized at the core and the perimeter of the skyrmion. As in Ref.~\cite{Yang2016}, we keep the total radial skyrmion size fixed and allow for multiple radial oscillations of the out-of-plane spin component, producing concentric annular regions with alternating polarity. 

The magnetization profile is parameterized as
\begin{equation}
    \mathbf{N}(r,\theta)
    = \bigl( \sin f(r)\cos(n_{\mathrm{sk}}\theta),\;
              \sin f(r)\sin(n_{\mathrm{sk}}\theta),\;
              \cos f(r) \bigr),
    \label{eq:N}
\end{equation}
where $(r,\theta)$ are polar coordinates about the skyrmion center, $n_{\mathrm{sk}}$ ($n_{\mathrm{sk}} = 1$ in this work) denotes the azimuthal winding number, and $f(r)$ controls the radial profile of the out-of-plane component.

We choose
\begin{equation}
f(r) =
\begin{cases}
    0, & 0 \le r < r_0, \\[4pt]
    \displaystyle \frac{\pi p\, (r - r_0)}{R - r_0}, & r_0 \le r < R, \\[6pt]
    \pi p, & r \ge R,
\end{cases}
\label{eq:f}
\end{equation}
where $R$ is the skyrmion radius and $p$, an odd integer, counts the number of radial oscillations of $\cos f(r)$ between $+1$ and $-1$. $p$ is thus effectively a measure of the induced spin-orbit coupling. A representative $p=1$ skyrmion configuration is shown in Fig.~\ref{fig:setup}.

\begin{figure}
\centering
\includegraphics[width=3.1in]{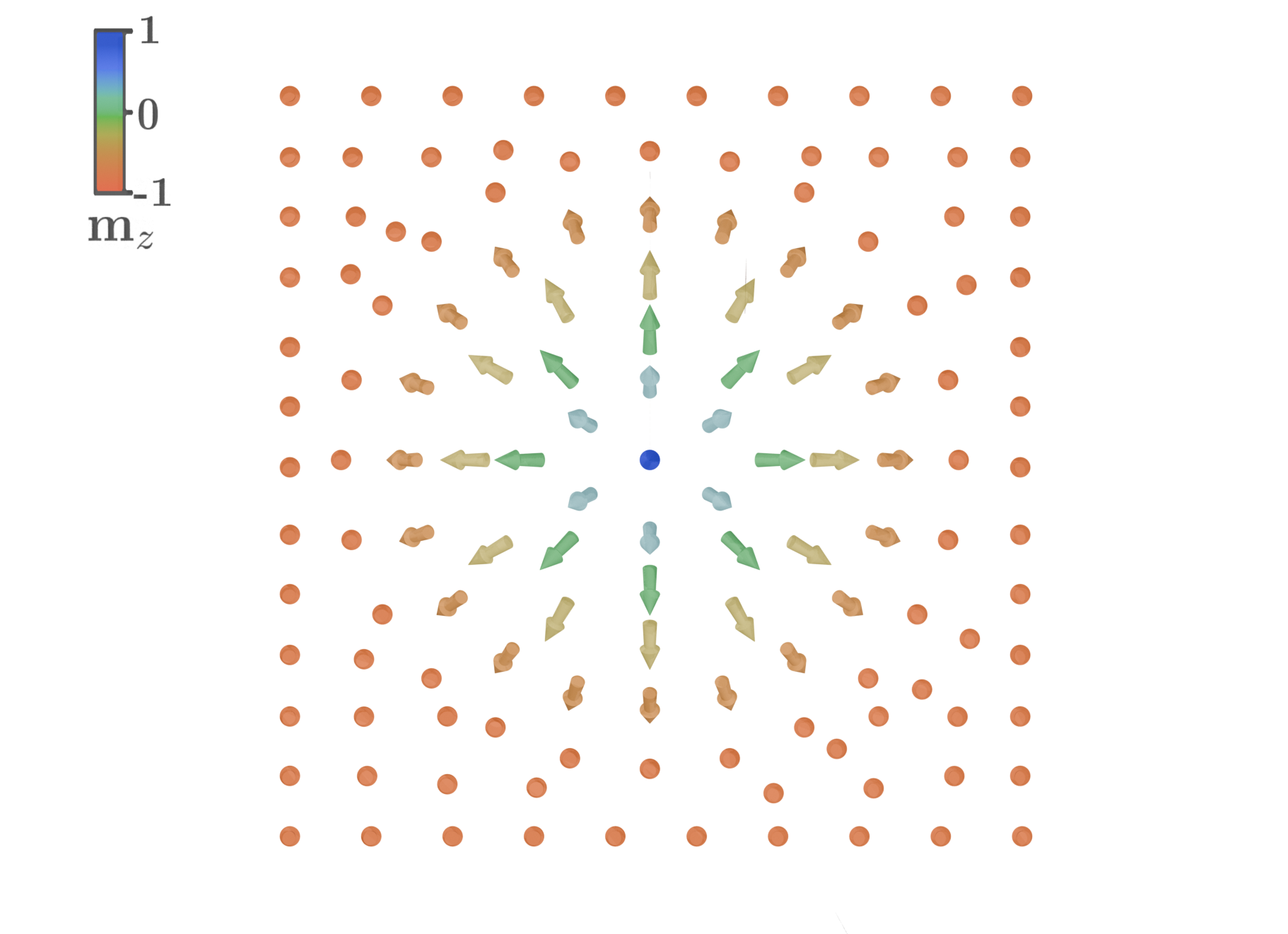}
\caption{Magnetization profile $\mathbf{N}(r,\theta)$ for a N\'eel skyrmion with $p=1$.}
\label{fig:setup}
\end{figure}

\subsection{Kinetic and pairing Hamiltonian}

We assume that the length scales associated with the skyrmion winding are relatively smooth compared to relevant length scales in the superconducting layer. The kinetic part of the superconducting electrons is then given by
\begin{equation}
    H_0 =
    \sum_{\beta}
    \int d^2\mathbf{r}\;
    \psi^\dagger_\beta(\mathbf{r})
    \left( -\frac{\nabla^2}{2m} - \mu \right)
    \psi_\beta(\mathbf{r}),
\end{equation}
where $\psi_\beta(\mathbf{r})$ annihilates an electron of spin $\beta=\uparrow,\downarrow$ at position $\mathbf{r}$.

We consider two fully gapped pairing states: a $d+is$ state and a chiral $d+id$ state. The corresponding pairing Hamiltonians 
in the presence of a vortex with winding $n_v$ are
\begin{align}
H_{\Delta_1} &= \sum_{\beta\gamma}
        \int d^2\mathbf{r}\;
        \psi^\dagger_\beta(\mathbf{r})\,
        e^{i n_v\theta}
        \Bigl[\Delta_d(\partial_x^2 - \partial_y^2)
        + i\Delta_s\Bigr]  \nonumber \\
        & \hspace{2cm}  \cdot (i\sigma^y_{\beta\gamma})
        \psi^\dagger_\gamma(\mathbf{r})
        + \text{H.c.}, \label{eq:pairing1} 
\\[4pt]
    H_{\Delta_2}
    &= \sum_{\beta\gamma}
        \int d^2\mathbf{r}\;
        \psi^\dagger_\beta(\mathbf{r})\,
        e^{i n_v\theta}
        \Bigl[\Delta_d(\partial_x^2 - \partial_y^2
        + 2 i\partial_x\partial_y)\Bigr]  \nonumber \\
         & \hspace{2cm} \cdot (i\sigma^y_{\beta\gamma})
        \psi^\dagger_\gamma(\mathbf{r})
        + \text{H.c.}, \label{eq:pairing2}
\end{align}
where $n_v$ is the vorticity, $\Delta_s$ and $\Delta_d$ are pairing amplitudes, and the factor of $\sigma^y$ enforces spin-singlet pairing.

\subsection{Skyrmion–electron coupling}

The conduction electrons couple to the skyrmion magnetization via a local exchange term
\begin{equation}
    H_{\mathrm{ex}} =
    \alpha
    \sum_{\beta\gamma}
    \int d^2\mathbf{r}\;
    \psi^\dagger_\beta(\mathbf{r})
    \bigl( \mathbf{N}(\mathbf{r})\cdot\boldsymbol{\sigma}_{\beta\gamma} \bigr)
    \psi_\gamma(\mathbf{r}),
\end{equation}
where $\alpha$ is the exchange coupling and $\boldsymbol{\sigma}$ are Pauli matrices in spin space. We focus on the physically relevant case $n_v=1$ throughout.

\subsection{Nambu representation}

Introducing the Nambu spinor
\begin{equation}
    \Psi(\mathbf{r})
    = \bigl(
        \psi_\uparrow^\dagger,
        \psi_\downarrow^\dagger,
        \psi_\downarrow,
        -\psi_\uparrow
      \bigr),
\end{equation}
the full Hamiltonian can be written compactly as
\begin{equation}
    H = \int d^2\mathbf{r}\;
    \Psi^\dagger(\mathbf{r})
    \left( h_k + h_j \right)
    \Psi(\mathbf{r}),
\end{equation}
where $h_k$ contains the kinetic terms and exchange coupling, and $h_1$ and $h_2$ correspond to the pairing Hamiltonians in Eqs.~\ref{eq:pairing1} and \ref{eq:pairing2}, respectively. Explicitly,
\begin{align}
    h_k
    &= \left(-\frac{\nabla^2}{2 m} - \mu\right)\tau^z + \alpha\,\mathbf{N}\cdot\boldsymbol{\sigma} ,
\\[4pt]
    h_1
    &= 
    - \Delta_s\bigl[\cos(n_v\theta)\,\tau^y
                      + \sin(n_v\theta)\,\tau^x\bigr]
    \nonumber\\
    &\quad
    + \Delta_d\cos(n_v\theta)(\partial_x^2 - \partial_y^2)\tau^x
    - \Delta_d\sin(n_v\theta)(\partial_x^2 - \partial_y^2)\tau^y,
\\[4pt]
    h_2
    &= \Delta_d\cos(n_v\theta)(\partial_x^2 - \partial_y^2)\tau^x
    - \Delta_d\sin(n_v\theta)(\partial_x^2 - \partial_y^2)\tau^y
    \nonumber\\
    &\quad
    - 2\Delta_d\cos(n_v\theta)\,\partial_x\partial_y\,\tau^y
    - 2\Delta_d\sin(n_v\theta)\,\partial_x\partial_y\,\tau^x.
\end{align}

Here $\tau^i$ are Pauli matrices in particle–hole space, while $\sigma^i$ act in spin space. 

\subsection{Topological winding condition}

A skyrmion with azimuthal winding $n_{\mathrm{sk}}$ and radial oscillation number $p$ generates an effective spin–orbit coupling whose strength grows with $p$; experimentally accessible skyrmions typically have $n_{\mathrm{sk}}=1$ and small $p$. As shown in Ref.~\cite{Rex2019}, binding a superconducting vortex to the skyrmion can compensate for the odd winding and enable a topological phase even when $n_{\mathrm{sk}}=1$. For an $s$-wave superconductor, MZMs appear when
\begin{equation}
    n_{\mathrm{tot}} = n_{\mathrm{sk}} + n_v \in 2\mathbb{Z},
    \label{eq:parity_swave}
\end{equation}
that is, when the total winding is even. 

For more general pairing symmetries, the pairing itself contributes angular momentum $n_{\mathrm{pair}}$ (e.g., $n_{\mathrm{pair}}=0$ for $s$-wave, $1$ for chiral $p$-wave, $2$ for chiral $d$-wave), analogously to the contribution from winding of the superconducting phase around the vortex. The topological criterion straightforwardly generalizes to 
\begin{equation}
    n_{\mathrm{tot}}
    = n_{\mathrm{pair}} + n_{\mathrm{sk}} + n_v
    \in 2\mathbb{Z}.
    \label{eq:parity_general}
\end{equation}

The above result can be rationalized as follows. The Majorana bound state pinned to a vortex can be realized whenever the solutions of the BdG equation allow for wavefunctions that transform with total angular momentum $l_z = 0$; the subspace of states with $l_z = 0$ is particle-hole symmetric which is a prerequisite to realize MZMs. A quasiparticle traversing around a vortex picks up a Berry phase of $\pi n_{\mathrm{sk}}$ due to the skyrmion texture besides a $\pi$ phase from both $n_{v}$ (due to real space winding of the superconducting order parameter) and $n_{\mathrm{pair}}$ (due to momentum space structure of the superconducting order parameter, which leads to a winding in real space as the particle traverses around the vortex with a momentum that is azimuthal). These phases must add up to a multiple of $2\pi$ to allow for a solution with $l_z = 0$. For example, a chiral p-wave superconductor naturally hosts MZMs bound to a vortex ($n_{\mathrm{sk}} = n_{v} = 1$).  

In what follows, since $d$ and $s$ have even $n_{\mathrm{pair}}$, we consider a heterostructure consisting of either a $d+is$ ($n_{\mathrm{pair}}=0$) or $d+id$ superconductor ($n_{\mathrm{pair}}=2$), pierced by a vortex with $n_v=1$ and coupled to a magnetic layer containing a skyrmion with $n_{\mathrm{sk}}=1$. This yields an even total winding and therefore allows for the emergence of MZMs.

\begin{figure}
\centering
\includegraphics[width=3.1in]{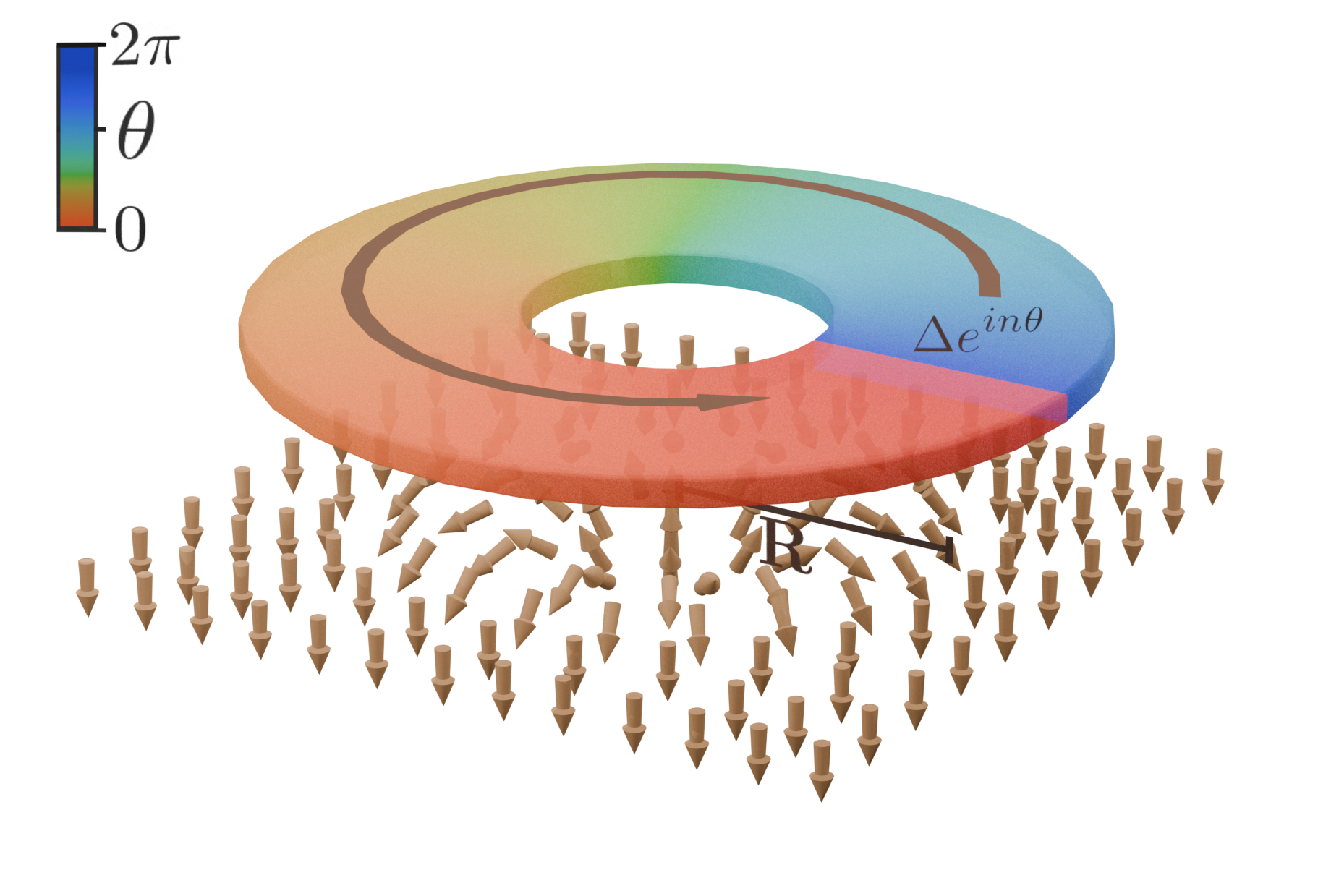}
\caption{Schematic heterostructure consisting of a $d+is$ or $d+id$ superconductor coupled to a magnetic layer hosting a N\'eel skyrmion with $p=1$.}
\label{fig:setup}
\end{figure}

\section{Numerical Simulations}\label{sec:Method}

\subsection{Discrete lattice model}\label{ssec:discrete}

To carry out numerical simulations we discretize the continuum Hamiltonians
introduced in Sec.~\ref{sec:Model} on a square lattice of spacing $a$. This lattice scale should not be confused with any intrinsic lattice scale in the superconductor.
As shown in App.~\ref{app:Discrete}, the discretization yields effective
lattice Hamiltonians $H_1$ and $H_2$ corresponding to the $d+is$ and $d+id$
pairing symmetries, respectively.  In terms of the Nambu spinor
$\Psi_{\mathbf r}$, the Hamiltonians take the form
\begin{equation}
    H_j = \sum_{\mathbf r, \mathbf r'} \Psi_{\mathbf r}^\dagger\, (\tilde{h}_k + \tilde{h}_j) (\mathbf r,\mathbf r')\,\Psi_{\mathbf r'} ,
    \label{eq:discrete_hamiltonians}
\end{equation}
with $j =1,2$ and 
\begin{align}
\tilde h_k
&= -\tau^z
    \Bigg[
        \frac{\delta_{\mathbf r,\mathbf r'+\hat x}
        + \delta_{\mathbf r,\mathbf r'-\hat x} + \delta_{\mathbf r,\mathbf r'+\hat y} + \delta_{\mathbf r,\mathbf r'-\hat y}}{2ma^2}
         \nonumber \\
        & \qquad \qquad \qquad + \Bigl(\mu - \frac{2}{ma^2}\Bigr)\delta_{\mathbf r,\mathbf r'}
    \Bigg]
    + \alpha\,\mathbf N_{\mathbf r}\cdot\boldsymbol{\sigma}
\nonumber\\
\tilde h_1
&=  \frac{\Delta_d}{a^2}\left( \tau^x 
    \cos(n_v\theta_{\mathbf r,\mathbf r'}) - \tau^y
    \sin(n_v\theta_{\mathbf r,\mathbf r'})\right)
    \nonumber\\
&\quad
   \times \left(
        \delta_{\mathbf r+\hat x,\mathbf r'} + \delta_{\mathbf r-\hat x,\mathbf r'}
        - \delta_{\mathbf r+\hat y,\mathbf r'} - \delta_{\mathbf r-\hat y,\mathbf r'}
    \right)
\nonumber\\
&\quad
- \Delta_s\bigl[
        \sin(n_v\theta_{\mathbf r,\mathbf r'})\,\tau^x
        + \cos(n_v\theta_{\mathbf r,\mathbf r'})\,\tau^y
    \bigr]\delta_{\mathbf r,\mathbf r'}
+ \mathrm{H.c.},
\nonumber \\[6pt]
\tilde h_2
&= \frac{\Delta_d}{a^2} \left(\tau^x 
    \cos(n_v\theta_{\mathbf r,\mathbf r'}) - \tau^y
    \sin(n_v\theta_{\mathbf r,\mathbf r'})\right)
    \nonumber\\
&\quad
   \times
    \left(
        \delta_{\mathbf r+\hat x,\mathbf r'} + \delta_{\mathbf r-\hat x,\mathbf r'}
        - \delta_{\mathbf r+\hat y,\mathbf r'} - \delta_{\mathbf r-\hat y,\mathbf r'}
    \right)
\nonumber\\
&\quad
- \frac{\Delta_d}{2a^2} \left(\tau^y 
    \cos(n_v\theta_{\mathbf r,\mathbf r'}) - \tau^x
    \sin(n_v\theta_{\mathbf r,\mathbf r'})\right)
    \nonumber\\
&\quad \times
    \left(
        \delta_{\mathbf r+\hat x+\hat y,\mathbf r'} + \delta_{\mathbf r-\hat x-\hat y,\mathbf r'}
        - \delta_{\mathbf r+\hat x-\hat y,\mathbf r'}
        - \delta_{\mathbf r-\hat x+\hat y,\mathbf r'}
    \right)
\nonumber\\
&\quad
+ \mathrm{H.c.}
\end{align}

Here $\theta_{\mathbf r,\mathbf r'} = (\theta_{\mathbf r} + \theta_{\mathbf r'})/2$ is the center-of-mass polar angle associated with the bond connecting $\mathbf r, \mathbf r'$. We measure distances in units of the lattice constant $a$ which is set to $1$.

\medskip
For the purposes of numerically studying the existence of MZMs in this setting, we work with an annual geometry wherein all sites satisfying $r < r_0$ or $r > R$ are discarded. Here $r_0$ parameterizes the vortex core size and $R$ corresponds to the skyrmion radius. Throughout the simulations we take $r_0 = 2a, R = 175.5\,a$. Note that it is the radial gradient of the winding of the skyrmion texture, $f'(r) \approx \pi p / R$, that sets the strength of the induced spin orbit coupling. The role of the skyrmion size $R$ here is to simply extend the region where such coupling is applicable to as far out as numerically feasible to obtain sharper phase diagrams. From an experimental point of view, any skyrmion width beyond the London screening length in the superconductor is not useful; we implicitly assume this screening length is larger than $R$. Likewise, any extent of the skyrmion within the vortex core is also not modeled as it does not engender spin-orbit coupling inside the core. 

Hereon, all energies are expressed in units of $1/(2ma^2) \equiv 1$ and all lengths in units of $a \equiv 1$. The energy scale $1/(2ma^2)$ should not be interpreted as setting the bandwidth of free electrons in the superconductor---importantly, it sets the scale of the Fermi wavevector $k_F$ and Fermi velocity $v_F$ in tandem with the strength of the exchange coupling $\alpha$ and the induced spin orbit effective Rashba spin-orbit coupling that, as we show, appears with strength $\beta = f'/4m$. In the semiclassical, Andreev approximation, these parameters effectively determine the phase diagram. 

For the $d+is$ simulations we use
$\Delta_s=1$ and $\alpha=1.5$, while $\Delta_d$ and $p$ are varied.  For $d+id$ we again set
$\alpha=1.5$ and treat $\Delta_d$ and $p$ as tunable parameters. We choose the values $\Delta$ to be as large as feasible to obtain a clear phase diagram, and $\alpha > \Delta$ is chosen to allow for the existence of the topological phase. (This is equivalent to the usual requirement that an in plane magnetic field must be larger than the induced gap to engender MZMs in a variety of spin-orbit coupled superconducting heterostructures.) For simplicity, we also choose $\mu = 0$. Note that at large $\alpha \gg \beta k_F$, and $\mu = 0$, the Fermi wave-vector $k_F$ is determined by $k^2_F/2m \approx \alpha$. This amounts to a Fermi wavelength $\lambda_F/a = 2\pi/\alpha \approx 4.2$ which is indeed several times greater than the lattice constant, justifying the validity of the discretization.  


\subsection{Identifying topological phases}
\label{sec:Identify}

We use exact sparse matrix diagonalization to compute the lowest few energy (by absolute value) eigenvalues and eigenstates of the Hamiltonians discussed above. To determine whether the system hosts MZMs on the inner and outer edges of the annulus, we first identify the two states $\pm E$ closest to zero energy.  In a topological phase, these states lie exponentially close to zero and are exponentially localized at opposite boundaries.  However, in two-dimensional systems, a continuum of states corresponding to edge mode(s) can appear inside the bulk gap at the edge for a variety of reasons---for instance, it is well known that for certain terminations of the lattice, a nodal d-wave superconductor can spontaneously exhibit edge currents that are not associated with chiral or helical Majorana modes~\cite{pathak2024edge}. Consequently, neither localization length nor energy alone provides a reliable signature of MZM physics. (Numerical illustration of these challenges is discussed in App.~\ref{app:edge_modes}.) We therefore extract the MZM wavefunctions directly by exploiting a key property of MZMs---that the sum and difference of the particle and hole components of the MZM wavefunction are localized at opposite edges. Note this contrasts with regular Andreev bound states which can appear localized as well, but only at a single edge; see Fig.~\ref{fig:probability} for example. 

\begin{figure}[t]
\centering
\includegraphics[width=0.5\textwidth]{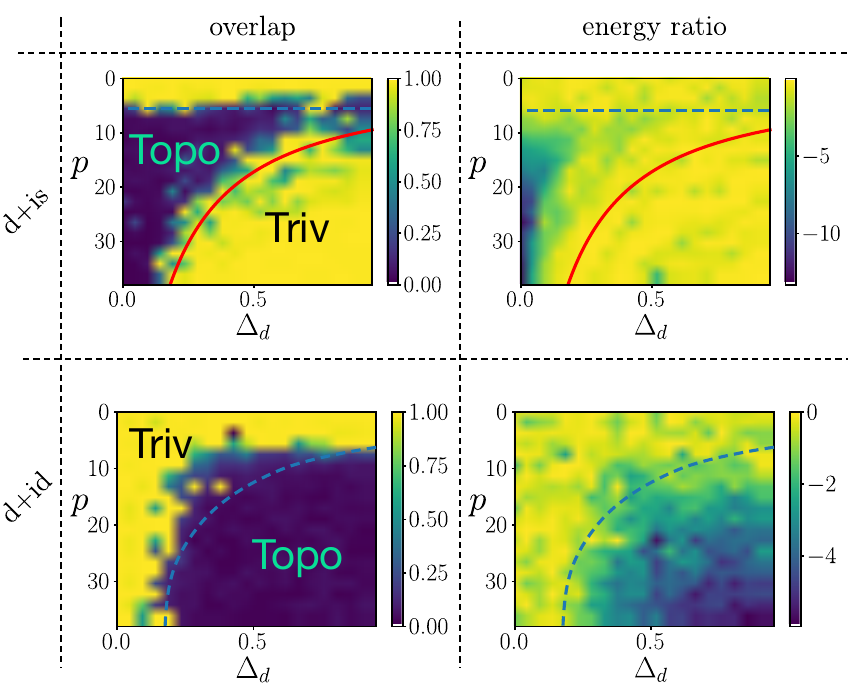}
\caption{Topological phase diagrams for the $d+is$ and $d+id$ skyrmion–superconductor models.
Left column: Majorana overlap criterion (low overlap = topological).
Right column: energy-ratio criterion. In all panels the inner and outer radii of the annulus are
$r_0=2a$ and $R=175.5a$, respectively. The red curve in the $d+is$ panel indicates a heuristic estimate of the phase boundary based on competition between singlet and skyrmion-induced triplet terms. The
blue dashed curve for the $d+id$ case is a transition arising 
from finite size
effects, as discussed in the text, with the topological phase
extending leftwards on larger system sizes.}
\label{fig:phase}
\end{figure}
Let $P$ denote the particle--hole conjugation operator, and let $\ket{\Psi_E}$ be the positive-energy eigenstate closest to zero.  In the absence of numerical phases, the two Majorana modes would be
\begin{align}
    \ket{\Psi_L} &\sim \ket{\Psi_E} + P\ket{\Psi_E}, \\
    \ket{\Psi_R} &\sim \ket{\Psi_E} - P\ket{\Psi_E}.
\end{align}
In practice, exact diagonalization returns the eigenvector only up to an arbitrary phase,
$\ket{\Psi'_E} = e^{i\phi}\ket{\Psi_E}$.  We therefore introduce a trial corrective phase
$e^{i\theta}$ and define
\begin{align}
    \ket{\Psi'_L}(\theta) &= e^{i\theta}\ket{\Psi'_E}
        + e^{-i\theta}P\ket{\Psi'_E}, \\
    \ket{\Psi'_R}(\theta) &= e^{i\theta}\ket{\Psi'_E}
        - e^{-i\theta}P\ket{\Psi'_E}.
\end{align}

For each $\theta$ we compute the radial probability distributions
$P^{\theta}_{L}(r)$ and $P^{\theta}_{R}(r)$, 
\[
    P^{\theta}_{L/R}(r) = \sum_{x,y \in \mathcal{X}_r} \left| \Psi_{L/R} (x,y)\right|^2.
\]
where $\left| \Psi_{L/R} (x,y)\right|^2$ is the total amplitude, summing over spin and particle--hole components, of the putative left (L) and right (R) MZM wavefunctions at  lattice point $(x,y)$. The amplitude at distance $r$ is computed by summing over points $(x,y)$ inside non-overlapping annular regions $\mathcal{X}_r$ centered at radial distance $r$ and width $2a$. We then evaluate their spatial overlap
\[
    I^\theta = \sum_r P^{\theta}_{L}(r)\,P^{\theta}_{R}(r).
\]
The optimal phase $\theta^\star$ is the one minimizing $I^\theta$, and the corresponding
$\ket{\Psi'_{1,2}}(\theta^\star)$ are identified as the physical Majorana modes. A phase is deemed topological whenever the optimal overlap satisfies
\[
    I^{\theta^\star} \ll 1,
\]

Scanning parameter space and evaluating $I^{\theta^\star}$ allows us to fairly unambiguously determine phase boundaries of the transition. For comparison, we also show color plots for the ratio of the second lowest positive energy eigenvalue $E_2$, over the lowest positive energy $E_2$, that is, $E_2/E_1$. Naively, one would expect that such a ratio is large for the topological phase but finite for the trivial phase. However, a direct comparison shows that this method does not yield clear phase boundaries; see Fig.~\ref{fig:phase}.

\subsection{Numerical results}\label{sec:Results}

\begin{figure}[t]
\centering
\includegraphics[width=\linewidth]{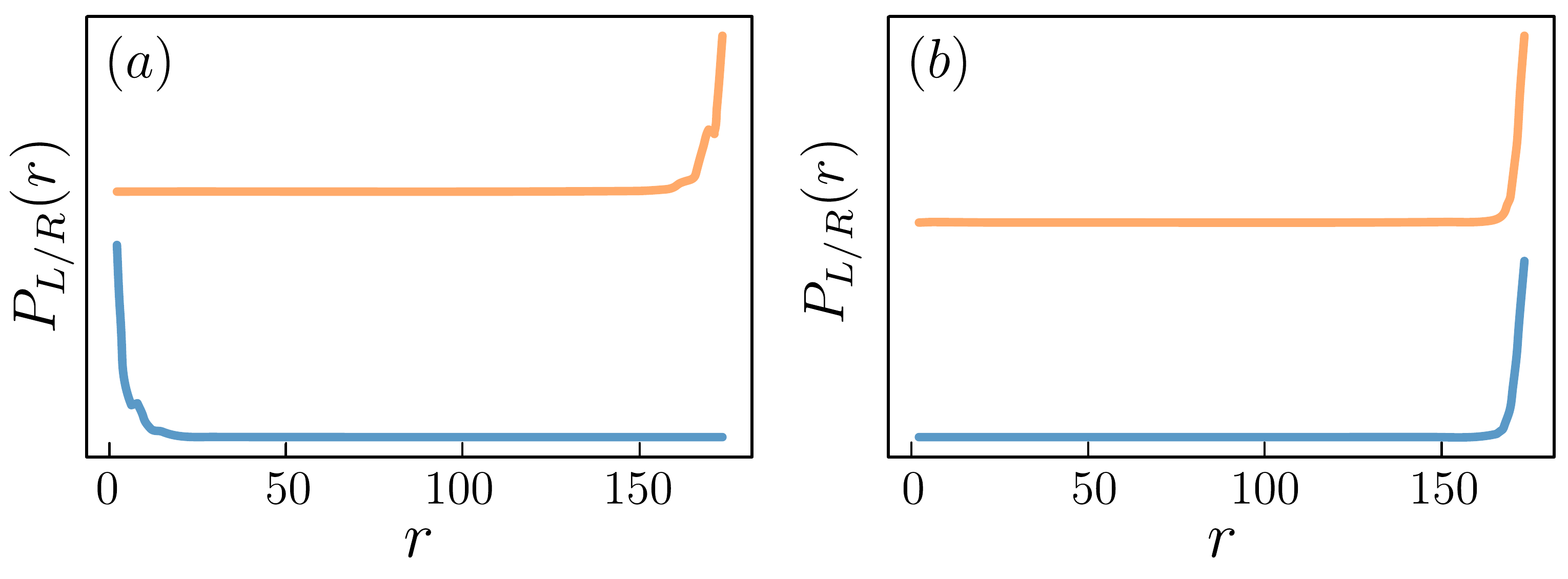}
\caption{Radial probability distributions of the two edge Majorana modes for the $d+is$ model.
(a) A point deep in the topological phase ($p=16$, $\Delta_d=0.1$) showing localization on opposite boundaries.
(b) A point in the trivial phase ($p=30$, $\Delta_d=0.75$), where both modes localize at the outer edge.} 
\label{fig:probability}
\end{figure}

Using the procedure outlined above, we compute the phase diagrams for both the $d+is$ and $d+id$ models.  Figure~\ref{fig:phase} captures our main finding: the first column shows the overlap criterion, while the second column uses the ratio of the lowest positive energy to the next-lowest energy as an alternative diagnostic.  Both criteria yield consistent phase boundaries, though the overlap method produces a sharper distinction between trivial and topological regions. Figure~\ref{fig:probability} shows representative radial probability distributions for two parameter points where the energy ratio produces a false positive for the topological phase by virtue of detecting a trivial Andreev bound state, while the MZM overlap does not.

For sufficiently small $p$ (the radial oscillation number controlling the effective spin--orbit coupling) in either pairing state, the system is always trivial---this is a finite size effect which occurs due to the divergence of the size of the MZMs in the limit of small spin orbit coupling. In the $d+id$ case, tuning $\Delta_d$ to $0$ also leads to the bulk gap closing, and leads to another finite size revival of the trivial phase because of the divergence of the size of MZMs. This behavior is identical to that seen in the pure $s$-wave case and reflects the need for sufficient skyrmion-induced spin--orbit coupling to stabilize MZMs.

Counterintuitively, we find that in the $d+is$ model, increasing either the superconducting strength $\Delta_d$ or the effective spin orbit strength $p$ eventually destroys the topological phase, in agreement with the analytical considerations of Sec.~\ref{sec:Analysis}.  In contrast, in the $d+id$ case, the topological phase remains stable at large $\Delta_d$ and $p$, as can be conventionally expected. We rationalize these findings  by analyzing the effect of the skyrmion in a modified basis wherein the effective magnetic field felt by the superconducting electrons is rotated to point uniformly in the $+z$ direction.

\section{Analysis}\label{sec:Analysis}

In this section we analyze the continuum Hamiltonian in the rotated basis defined by the skyrmion texture, and explain the mechanisms underlying the phase boundaries observed in Sec.~\ref{sec:Results}. 

\subsection{Hamiltonian in the rotated basis: SOC generation and transformation of the superconducting order}
\label{subsec:rotatedbasis}

To clarify how the magnetic texture affects both the normal-state and pairing sectors, we
apply the spatially varying spin rotation  
\begin{equation}
U(\mathbf{r})
    = \exp\!\left[
        \frac{i f(r)}{2}
        \bigl(
            \cos(n_{\mathrm{sk}}\theta)\,\sigma^y
            - \sin(n_{\mathrm{sk}}\theta)\,\sigma^x
        \bigr)
    \right],
    \label{eq:Unitary_recap}
\end{equation}
which aligns the spin quantization axis with the skyrmion field $\mathbf{N}(r,\theta)$.
As shown in Ref.~\cite{Yang} and summarized in App.~\ref{app:Unitary}, the transformed
kinetic term, for $n_{\text{sk}} = 1$, becomes  
\begin{align}
U h_k U^\dagger
&=
\left(
    -\frac{\nabla^2}{2m}
    - \mu
    - \frac{f'(r)^2}{8m}
\right)\tau^z
+ \alpha \sigma^z
\nonumber \\ &\quad
+ e^{-i\sigma^z\theta}
\left[
    \frac{i\sin f(r)}{4mr^2}\,\partial_\theta\,\sigma^x
    - \frac{i f'(r)}{4m}\,\partial_r\,\sigma^y
\right]\tau^z ,
\label{eq:rotated_kinetic}
\end{align}
where the $\alpha\sigma^z$ term represents a uniform Zeeman field in the rotated basis, and the last
line encodes an emergent, spatially varying spin–orbit coupling.  
This SOC resembles the Rashba structure but carries a nontrivial dependence on the radial
profile $f(r)$ and its derivative $f'(r)$. In particular, at large $r$, ignoring the first term $\sim 1/r^2$, and assuming constant $f'(r)$, the SOC becomes precisely of the Rashba form. 

Terms proportional to $\tau^z$ with weak position dependence also appear, corresponding to an
effective modulation of the chemical potential. These terms do not qualitatively effect the phase diagram.

For the pairing terms, the transformation behaves qualitatively differently depending on the
pairing symmetry. Since the $s$-wave pairing considered is spin singlet and momentum independent, it
commutes with $U$ and remains unchanged. In contrast, momentum-dependent order parameters—such as
the $d+is$ and $d+id$ pairings considered here—do \textit{not} commute with $U$.  
The derivatives act both on the fermion fields and on the spatially varying matrix
$U(\mathbf r)$, producing new contributions with angular dependence different from that of the original pairing channel. 

To capture the leading effects analytically, we retain only the terms that survive in the
limit where the vortex core radius and skyrmion radius are comparable and both are much larger
than microscopic length scales. Under this controlled approximation, the transformed
pairing terms can be expressed as the original singlet components supplemented by additional
triplet-like terms proportional to $f'$.  
These emergent triplet components play a central role in determining the topological phase
boundaries. In particular, we find the effective superconducting order as a function of space and momentum (with $\partial_{x(y)} \rightarrow i k_{x(y)}$) for the $d+is$ and $d+id$ cases to be, 
\begin{align}
\Delta_{\mathrm{eff}}^{(d+is)}
&=
\Big[
    \Delta_d(\partial_x^2 - \partial_y^2)
    + i\Delta_s
\Big] i\sigma^y
-
\frac{\Delta_d f'(r)^2}{4}\cos(2\theta)\, i\sigma^y
\nonumber \\[4pt] &\quad
-
\Delta_d f'(r)\,
e^{-i\theta\sigma^z}
\bigl[
    \cos\theta\,\partial_x
    - \sin\theta\,\partial_y
\bigr] ,
\label{eq:Delta_eff_dis}
\end{align}

\begin{align}
\Delta_{\mathrm{eff}}^{(d+id)}
&=
\Delta_d
\Big[
    (\partial_x^2 - \partial_y^2)
    + 2i\,\partial_x\partial_y
    - \tfrac{f'(r)^2}{4} e^{2i\theta}
\Big] i\sigma^y
\nonumber \\[4pt] &\quad
-
\Delta_d f'(r)\,
e^{-i\theta\sigma^z}
\bigl[
    \cos\theta\,(\partial_x - i\partial_y)
    - \sin\theta\,(\partial_y - i\partial_x)
\bigr] .
\label{eq:Delta_eff_did}
\end{align}

\subsection{Competition between angular-momentum sectors and the role of the projected order parameter $\Delta_{--}$}
\label{subsec:projection}

The additional pairing terms $\propto f'$ generated by the rotation $U$ carries angular momentum
that is odd and thus different from that of the bare $d$-wave order parameter. These terms behave effectively as induced \textit{spin-triplet} pairing, competing with
the original spin-singlet structure. This has the important 
consequence that when such terms become dominant in amplitude, the criterion for obtaining MZMs, Eq.~\ref{eq:parity_general}, is violated. Thus, we may expect the topological transition to occur when these terms become comparable in magnitude. We can do so by substituting \(\partial_r \to ik_F\). For the $d+is$ case, we obtain
\[
\Delta^{\rm singlet}_{d+is} \sim \max(\Delta_s,\;\Delta_d k_F^2,\; \Delta_d f'^2),
\qquad
\Delta^{\rm triplet}_{d+is} \sim \Delta_d f' k_F.
\]
Since $\Delta_s$ is angle-independent, it is the relevant singlet scale, giving the phase-boundary condition
\begin{equation}
    \frac{\Delta_s}{\Delta_d f' k_F} \sim \mathcal{O}(1),
\end{equation}
which explains the numerically observed curve,  $\frac{\Delta_s}{\Delta_d f' k_f} = 0.3$ (shown in red in Fig.~\ref{fig:phase}) which yields a transition for \(|f'| \propto 1/\Delta_d\), or \(p \propto 1/\Delta_d\). 

For the $d+id$ case, one may expect a similar transition. Here, 
\[
\Delta^{\rm singlet}_{d+id} \sim \Delta_d k_F^2, \qquad
\Delta^{\rm triplet}_{d+id} \sim \Delta_d f' k_F.
\]
Thus,
\begin{equation}
    \frac{\Delta^{\rm singlet}_{d+id}}{\Delta^{\rm triplet}_{d+id}} \sim \frac{k_F}{f'},
\end{equation}
and a transition can only occur in the extreme limit when the magnetic order of the skyrmion radially oscillates at a wave-vector close to $k_F$. 

A more accurate way to predict the boundary of this phase transition is to examine the effective superconducting order parameter projected into the lower band (split due to spin orbit coupling and applied field) and assess when it closes for any $\mathbf{r}, \mathbf{k}$, with $\mathbf{k}$ being a momentum on the Fermi surface. Let $\ket{u_\pm(\mathbf{k})}$ denote the eigenstates of the rotated normal-state Hamiltonian,
corresponding to the upper ($+$) and lower ($-$) spin–orbit–coupled bands. The effective pairing within the lower band is
\begin{equation}
    \Delta_{--}(\mathbf{k},\theta)
    =
    \bra{u_-(\mathbf{k})}
        \Delta_{\mathrm{eff}}(\mathbf{r},\mathbf{k})
    \ket{u_-(-\mathbf{k})},
\end{equation}
where $\Delta_{\mathrm{eff}}$ is the full, rotated superconducting order parameter.
By definition, $\Delta_{--}$ transforms as a chiral $p$-wave order parameter because
pairing within a single spin–orbit–split band necessarily carries odd angular momentum.
This is the channel responsible for generating MZMs. It receives contributions from both the spin-triplet and spin-singlet components in Eqs.~\ref{eq:Delta_eff_dis},~\ref{eq:Delta_eff_did}. 

To analyze the angular structure, we write the momentum as  
\[
\mathbf{k} = k_F(\cos\varphi_{\mathbf{k}},\,\sin\varphi_\mathbf{k}),
\]
with $\varphi_\mathbf{k}$ the polar angle in momentum space.  
The superconducting order $\Delta_{--}(\theta,\varphi_{\mathbf{k}})$ thus depends both on the real-space
angle $\theta$ (entering through $U(\mathbf r)$ and $f'(r)$) and the momentum-space angle
$\varphi_{\mathbf{k}}$ (entering through the form factors of the $d$-wave pairing).

Topological superconductivity requires that $\left|\Delta_{--}\right|$ is finite for all $\theta, \varphi_{\mathbf{k}}$. 
Equivalently, a phase transition occurs when there exists a pair $(\theta,\varphi_{\mathbf{k}})$ for which
\begin{equation}
    \Delta_{--}(\theta,\varphi_{\mathbf{k}}) = 0,
\end{equation}
that is, when the induced $p$-wave pairing amplitude vanishes on the Fermi surface for
some angular configuration.

It is easy to check this gap indeed closes at finite and large SOC $ \propto f'$ in the d+is case. Consider for simplicity the limit of large field $\alpha \sigma_z$ compared to the effective SOC which is, $\beta (k_x \sigma^y - k_y \sigma^x) $ where $\beta = f'/4m$.

To first order in $\delta = \beta k_F / \sqrt{\alpha^2 + (\beta k_F)^2}$, we find the projected pairing amplitude (see App.~\ref{app:projection_lower_band}) in the d+is case to be given by

\begin{align}
    &\Delta^{d+is}_{--}(\mathbf{k}, \theta) = i e^{i \phi_k} \delta \left[ - \Delta_d k^2_F \cos 2 \phi_k + i \Delta_s \right] \nonumber \\
    &-  e^{-i\phi_k} \delta \frac{\Delta_d f'^2}{4} \cos 2 \theta + i \Delta_d f' k_F e^{i\theta} \cos (\theta + \phi_k),
\end{align}

where the first term arises from the usual $d+is$ term in a Rashba spin-orbit coupled system and the last term corresponds to a triplet pairing contribution noted above. Invoking semi-classical arguments, we expect the quasiparticle travels with azimuthal angle $\phi_k = \theta + \pi/2$ around the vortex. It is easy to then check that at $\theta = \pi/4, 3\pi/4, $ etc., and $\phi_k = \theta + \pi/2$, only the s-wave term along with the triplet pairing term contributes, and the gap closes for $\delta \Delta_s = \Delta_d f' k_F$. 

A similar analysis in the d+id case immediately suggests that any topological transition at large $p$ is unlikely or occurs in a highly fine tuned way given that the parent d-wave gap never closes at any $\varphi_{\mathbf{k}}$. Note that in the case of extended $s$-wave pairing, one would generically also obtain additional spin-triplet terms, but similarly to the $d+id$ case, such terms would be much weaker. 





\subsection{Finite-size–driven transitions at small $p$ and small $\Delta_d$}
\label{subsec:finitesize}

The phase diagrams of Sec.~\ref{sec:Results} reveal an additional transition to a trivial
phase when both $p$ and $\Delta_d$ are small. This boundary is unrelated to the competition
between singlet and triplet components discussed above. Instead, it arises from a finite-size
effect associated with the localization length $\zeta$ of the Majorana modes.

The localization length is controlled by the effective $p$-wave gap induced in the lower band
and scales approximately as  
\[
\zeta \sim \frac{v_F}{E_\Delta},
\]
where $v_F$ is the Fermi velocity and $E_\Delta$ is the induced gap in the projected band.
For small $p$, or $\alpha/\beta k_F \gg 1$, the emergent SOC is weak, and hence $E_\Delta \approx \Delta \beta k_F / \alpha$ becomes small. Once $\zeta$
exceeds the system radius $R$, the two edge Majorana modes hybridize strongly and the system
becomes gapped. Plugging in $\zeta = R$ yields the condition $ p = 4 \alpha / \Delta$. For the $d+is$ case, we can ignore the smaller $d$-wave contribution, finding the transition to be at $p \approx 6$, which agrees well the finite size transition see in Fig.~\ref{fig:phase} for $d+is$. For the $d+id$ case, the condition $p = 4 /\alpha \Delta_d$ now represents a hyerpobola in the $p-\Delta_d$ plane which is in reasonably good agreement with the finite-size transition seen in Fig.~\ref{fig:phase}.

\section{Conclusion}

In this work we have investigated the emergence of Majorana zero modes in two-dimensional superconductors with unconventional pairing symmetries coupled to a chiral magnetic texture hosting a N\'eel-type skyrmion. Building on earlier results for conventional $s$-wave systems, we have shown that the skyrmion texture serves as an effective generator of both a spatially varying spin--orbit coupling and a uniform Zeeman field once transformed into a local spin-aligned basis. This mechanism enables topological superconductivity without the need for intrinsic spin--orbit coupling or externally applied magnetic fields, thereby providing an alternative platform for realizing MZMs.

A central finding of our study is that the presence of momentum-dependent pairing fundamentally modifies the effect of the skyrmion-induced transformation. For $d$-wave--based superconductors, the spatial gradients of the unitary rotation generate additional pairing components not present in the original Hamiltonian. These induced terms can be interpreted as spatially anisotropic triplet contributions, which compete with the intrinsic singlet sector and can either enhance or suppress topological superconductivity depending on microscopic parameters. Our analysis shows that the resulting topology depends sensitively on the interplay between the skyrmion winding number, the radial oscillation parameter $p$, and the strength of the $d$-wave component $\Delta_d$.

Using exact diagonalization of a discretized annular geometry, we constructed phase diagrams for both the $d+is$ and $d+id$ scenarios. In each case we employed two independent diagnostics---the overlap between inner and outer edge Majoranas and the low-energy spectral structure---to identify the topological regime. For the $d+is$ system, we observed that sufficiently large $\Delta_d$ or large skyrmion oscillation number $p$ destabilizes the topological phase by enhancing the induced triplet sector. In contrast, the $d+id$ system exhibits a qualitatively different behavior: although triplet components are also generated, they remain comparatively weak over a broad parameter range, allowing the topological phase to persist even at large $\Delta_d$.

We further provided an analytic framework explaining these numerical trends. By projecting the transformed pairing terms into the spin--orbit--coupled band and identifying the effective $\Delta_{--}$ component controlling the topological gap, we argued that the transition occurs when the triplet and singlet amplitudes become comparable. This reasoning naturally yields phase boundaries consistent with our simulations and clarifies why different pairing symmetries respond differently to the skyrmion-induced transformation. 

Overall, our results demonstrate that skyrmion-based engineering of topological superconductivity remains viable in systems with unconventional pairing, but the detailed topology is controlled by how the spatially varying spin rotation interacts with the momentum structure of the order parameter. These findings broaden the landscape of materials and heterostructures capable of hosting MZMs, especially in light of recent advances in stabilizing and electrically controlling N\'eel skyrmions in thin-film magnets. They also suggest that unconventional superconductors---already believed to host rich internal structures---may exhibit new forms of topological behavior when interfaced with chiral magnetic textures.

Turning to the potential for experimental realizations of our proposal, the two important ingredients, namely, a topological skyrmion 
texture and a superconductor, can be realized using a heterostructure 
formed by a chiral magnet and a type-II superconducting layer. In this geometry a very weak magnetic field can be sufficient to induce a skyrmion
in the magnetic layer and a vortex in the superconductor which can form a bound pair. Indeed, such heterostructure based skyrmion-vortex bound states 
have been previously found in conventional $s$-wave superconductors such as Nb coupled to magnetic multilayers of [IrFeCoPt] which host
strong spin-orbit coupling and chiral Dzyaloshinskii-Moriya interactions \cite{Petrovic2021}. In addition, there are a wide range of known
skyrmion materials including MnSi \cite{Muhlbauer_2009},
Cu$_2$OSeO$_3$ \cite{White_2012}, and Fe$_{1-x}$Co$_x$Si\cite{Munzer_2010,Yu_2010}, as well as more recently
discovered examples with small and more complex skyrmion textures including Gd$_2$PdSi$_3$ \cite{Kurumaji_2019}, 
GdRu$_2$Si$_2$ \cite{Hirschberger_2019} and Fe$_3$GeTe$_2$ \cite{Ding_2020}.
Since our proposal envisions $d+id$ or $d+is$ 
superconductors, a possible route is to explore candidate materials such as twisted cuprate layers \cite{ZhaoTwisted2021,ZhaoTRSB2023} or 
magnetically driven Cooper pairing in triangular geometries \cite{Ming2023chiralSC,wu2025microscopicfingerprintchiralsuperconductivity,Kim2023triangular}
which have been shown to host such time-reversal broken symmetry states. Remarkably, the cuprate superconductors have also been proposed to 
intrinsically host $d+is$ superconductivity given observations of a nodeless gap and a nonzero polar Kerr effect in the highly underdoped regime  \cite{Razzoli2013nodeless,Karapetyan2014Kerr,Mallik2020chiraldwave}. It would be interesting in the future to also explore new materials with
chiral exchange interactions and pairing which
might intrinsically and simultaneously host both chiral skyrmion textures and chiral superconductivity without the need to create
heterostructures.

On the theoretical front,
future work may focus on refining the continuum description, analyzing the effects of disorder and skyrmion motion, and identifying experimentally accessible material platforms where the required parameter regime can be realized. The ability to manipulate skyrmions electrically or magnetically opens promising avenues for dynamically tuning the topological phase and, potentially, for performing non-Abelian operations in skyrmion--superconductor hybrids.

\section{Acknolwedgments}

KA acknowledges support by the US Department of Energy, Office of Science, Basic Energy Sciences, Materials Sciences and Engineering Division. TP-B acknowledges support from ICFO for their hospitality while this work was being carried out. AP was supported by NSERC of Canada via Discovery grant RGPIN-2021-03214.  

\appendix
\section{Unitary Transformation and Transformed Hamiltonian}\label{app:Unitary}

In this appendix, we explicitly derive the transformed Hamiltonian under the local spin rotation defined in Eq.~\eqref{eq:Unitary_recap}.  
The untransformed single-particle Hamiltonian reads
\begin{equation}
    h_k = \left(-\frac{\nabla^2}{2m} - \mu\right)\tau^z + \alpha\,\mathbf{N}(r,\theta)\!\cdot\!\boldsymbol{\sigma},
\end{equation}
where $\mathbf{N}(r,\theta)$ is the spatially varying N\'eel skyrmion texture.

We apply the unitary transformation
\begin{equation}
    U(\mathbf{r}) = \exp\!\left[i\frac{f(r)}{2}\bigl(\cos\theta\,\sigma^y - \sin\theta\,\sigma^x\bigr)\right],
\end{equation}
which rotates the local spin quantization axis so that
\begin{equation}
    U(\mathbf{r})\,[\mathbf{N}(r,\theta)\!\cdot\!\boldsymbol{\sigma}]\,U^\dagger(\mathbf{r}) = \sigma^z.
\end{equation}
Thus, the exchange field becomes a uniform Zeeman term $\alpha \sigma^z$ in the rotated basis. 

Since $[-\mu\tau^z, U] = 0$, the only nontrivial contribution arises from transforming the kinetic term:
\begin{equation}
    U\left(-\frac{\nabla^2}{2m}\right)\tau^z U^\dagger
    = -\frac{\tau^z}{2m}
    \Bigl[
        \nabla^2 + 2(\nabla U^\dagger)\!\cdot\!\nabla + (\nabla^2 U^\dagger)
    \Bigr].
\end{equation}
It is convenient to evaluate the gradient in polar coordinates, 
\begin{equation}
    \nabla = \hat{\mathbf{e}}_r\,\partial_r + \hat{\mathbf{e}}_\theta\,\frac{1}{r}\partial_\theta,
\end{equation}
so that
\begin{align}
    (\nabla U^\dagger)\!\cdot\!\nabla 
    &= (\partial_r U^\dagger)\partial_r + \frac{1}{r}\,(\partial_\theta U^\dagger)\partial_\theta,\\
    \nabla^2 U^\dagger
    &= \partial_r^2 U^\dagger + \frac{1}{r}\partial_r U^\dagger + \frac{1}{r^2}\partial_\theta^2 U^\dagger.
\end{align}

After evaluating $\partial_r U^\dagger$ and $\partial_\theta U^\dagger$ and keeping only the symmetry-relevant contributions (see full derivation above), the transformed Hamiltonian becomes
\begin{align}\label{eq:AppTransformed}
    U h_k U^\dagger
    &=
    \left(-\frac{\nabla^2}{2m} - \mu - \frac{f'(r)^2}{8m}\right)\tau^z
    + \alpha\,\sigma^z
    \nonumber\\
    &\quad
    +\, e^{-i\sigma^z\theta}
    \left[
        \frac{i\sin f(r)}{4mr^2}\,\partial_\theta\,\sigma^x
        - \frac{i f'(r)}{4m}\,\partial_r\,\sigma^y
    \right]
    \tau^z.
\end{align}

The last line of Eq.~\eqref{eq:AppTransformed} constitutes an emergent, spatially dependent spin–orbit coupling generated entirely by the skyrmion texture. The term $-\frac{f'(r)^2}{8m}\tau^z$ corresponds to a weak spatially varying chemical potential shift, which is retained in the numerics but omitted in most analytical expressions.

\section{Discretized Tight-Binding Form}\label{app:Discrete}

We next discretize the full Hamiltonian on a square lattice of spacing \(a\), where $(i,j)$ labels lattice sites at position $\mathbf{r}_{ij} = (ia,ja)$. The continuum kinetic term is replaced by the standard second-order finite difference approximation:
\begin{align}
    \partial_x^2 \psi(x,y) &\rightarrow \frac{\psi_{i+1,j} - 2\psi_{i,j} + \psi_{i-1,j}}{a^2},\\
    \partial_y^2 \psi(x,y) &\rightarrow \frac{\psi_{i,j+1} - 2\psi_{i,j} + \psi_{i,j-1}}{a^2}.
\end{align}
Thus the discretized kinetic Hamiltonian is
\begin{align}
    H_t
    &=
    \sum_{i,j}
    \Biggl[
        -\frac{1}{2ma^2}
        \left(
            \Psi^\dagger_{i+1,j}\tau^z\Psi_{i,j}
            + \Psi^\dagger_{i,j+1}\tau^z\Psi_{i,j}
        \right)
        \nonumber\\
        &\quad
        + \left(
            -\mu + \frac{2}{ma^2}
        \right)
        \Psi^\dagger_{i,j}\tau^z\Psi_{i,j}
    \Biggr]
    + \mathrm{H.c.}
\end{align}

For the superconducting terms, derivatives are discretized analogously, while phases containing $\theta(\mathbf{r})$ are link-averaged to maintain gauge-compatible vortex winding (see Ref.~\cite{vafek2001quasiparticles} for more details):
\begin{equation}
    \theta^{ij}_{i'j'} = \frac{1}{2}\left[\theta(x_i,y_j)+\theta(x_{i'},y_{j'})\right].
\end{equation}

The $d$-wave nearest-neighbor contribution becomes
\begin{align}
    &H_{\Delta_d}
    =
    \frac{\Delta_d}{a^2}
    \sum_{i,j}
    \Bigl[
        \Psi^\dagger_{i+1,j}
        \bigl(
            \cos(n_v\theta^{i+1j}_{ij})\,\tau^x
            - \sin(n_v\theta^{i+1j}_{ij})\,\tau^y
        \bigr)
        \Psi_{i,j}
        \nonumber\\
        &\qquad
        - \Psi^\dagger_{i,j+1}
        \bigl(
            \cos(n_v\theta_{ij+1,ij})\,\tau^x
            - \sin(n_v\theta_{ij+1,ij})\,\tau^y
        \bigr)
        \Psi_{i,j}
    \Bigr]  \nonumber\\
    &\qquad + \mathrm{H.c.}
    \label{eq:dwavediscrete}
\end{align}

For the $d+id$ case, we encounter a mixed derivative term $\propto \partial_x \partial_y$. This extra contribution [compared to the d-wave case in Eq.~(\ref{eq:dwavediscrete})] is given by
\begin{align}
    H_{d+id}^{(xy)}
    &=
    \frac{\Delta_{d2}}{4a^2}
    \sum_{i,j}
    \Psi^\dagger_{i,j}\tau^y
    \bigl(
        \Psi_{i+1,j+1} + \Psi_{i-1,j-1}
        - \Psi_{i+1,j-1} \nonumber \\
        &- \Psi_{i-1,j+1}
    \bigr) + \mathrm{H.c.}
\end{align}

\section{Analytical estimate of phase boundaries}\label{app:PhaseLines}

In this appendix we present a detailed analytical derivation of the approximate phase boundaries obtained numerically in Sec.~\ref{sec:Analysis}. Our goal is to understand how the competition between the induced effective spin--triplet and the original spin--singlet pairing channels determines the transition between the topological and trivial phases.

We proceed using the following three–step program:
\begin{enumerate}
    \item Apply the local spin rotation $U(\mathbf{r}) = \exp\!\left[i\frac{f(r)}{2}\bigl(\cos\theta\,\sigma^y - \sin\theta\,\sigma^x\bigr)\right]$ to the superconducting pairing Hamiltonians corresponding to $d+is$ and $d+id$ order parameters.
    \item Focus on the bulk where all terms proportional to $1/r$ and $1/r^2$ may be safely discarded.
    \item Separate the transformed pairing into spin-singlet and spin-triplet sectors and compare their amplitudes to extract conditions for the breakdown of topological superconductivity.
    \item Project into the lower spin-orbit coupled band and check that the effective gap in this band obtains contributions from both terms, and closes at the observed phase boundary. 
\end{enumerate}

\subsection{Starting continuum pairing Hamiltonians}

The unrotated pairing Hamiltonians are
\begin{align}
    H^{(d+is)}_\Delta &= \left[ \Delta_d\,(\partial_x^2 - \partial_y^2)\, +i \Delta_s\, \right] (i \sigma^y),
    \label{eq:H_dis_appendix}\\
    H^{(d+id)}_\Delta &= \Delta_d\left[(\partial_x^2 - \partial_y^2)\, + 2i\,\partial_x\partial_y\,\right](i \sigma^y).
    \label{eq:H_did_appendix}
\end{align}
To perform the transformation, we express derivatives in polar form:
\begin{align}
    \partial_x &= \cos\theta\,\partial_r - \frac{\sin\theta}{r}\partial_\theta, \\
    \partial_y &= \sin\theta\,\partial_r + \frac{\cos\theta}{r}\partial_\theta.
    \label{eq:partialxandy}
\end{align}

First and second order spatial derivatives acting on $U^\dagger(\mathbf{r})$ generate new terms, including spin-triplet contributions.

\subsection{Large-core approximation}

To reduce complexity while retaining the leading-order spatial structure responsible for topological transitions, we focus on $r \gg \xi, 1/k_F$. The derivative $\partial_r U^\dagger_r$ is proportional to $f'(r)$ which is a constant $\propto p$. Thus, the action of the radial derivative in Eqs.~(\ref{eq:partialxandy}) remains finite while the azimuthal angle derivative dies at large $r$. Outside of finite size effects, the topological phase boundary is predicted by the bulk structure of the pairing and spin orbit coupling, thus we can safely discard terms that scale a s\(1/r\) or \(1/r^2\) . 


Operationally, this corresponds to dropping all \(\partial_\theta\)-dependent contributions as well as all terms arising explicitly from \(1/r\) and \(1/r^2\). Under this approximation,
\begin{equation}
    \partial_x \approx \cos\theta\,\partial_r,\qquad
    \partial_y \approx \sin\theta\,\partial_r,
\end{equation}
and therefore
\begin{align}
    \partial_x^2 - \partial_y^2 &\approx \cos(2\theta)\,\partial_r^2, \\
    \partial_x \partial_y &\approx \frac{1}{2}\sin(2\theta)\,\partial_r^2.
\end{align}

The following relations are helpful in obtaining the final form of the transformed superconducting term: 

\begin{align}
   \partial_r U^\dagger &= -\frac{f'(r)}{2}\sin\!\left(\frac{f(r)}{2}\right)
   - \frac{i f'(r)}{2}\cos\!\left(\frac{f(r)}{2}\right) \nonumber \\
   &\times 
   \bigl[\cos\theta\,\sigma^y - \sin\theta\,\sigma^x\bigr], \\
   \partial_r^2 U^\dagger &= -\frac{f'(r)^2}{4}\cos\!\left(\frac{f(r)}{2}\right)
   + \frac{i f'(r)^2}{4}\sin\!\left(\frac{f(r)}{2}\right) \nonumber \\
   &\times
   \bigl[\cos\theta\,\sigma^y - \sin\theta\,\sigma^x\bigr].
   \end{align} 

\subsection{Transformed pairing and decomposition into singlet/triplet}

After performing $U H_\Delta U^\dagger$ and collecting terms, we find the effective pairing operators
\begin{align}
    \Delta^{d+is}_{\rm eff}
    &=
    \Bigl[
        \Delta_d (\partial_x^2 - \partial_y^2) 
        + i\Delta_s
        - \frac{\Delta_d f'^2(r)}{4} \cos(2\theta)
    \Bigr]
        (i\sigma^y)
    \nonumber\\
    &\quad
    - \Delta_d f'(r)\,
    e^{-i\sigma^z\theta}\,
    \bigl[\cos\theta\,\partial_x - \sin\theta\,\partial_y\bigr],
    \label{eq:EffPair_is_final}
\\[4pt]
    \Delta^{d+id}_{\rm eff}
    &=
    \Delta_d\Bigl[(\partial_x^2 - \partial_y^2) + 2i\,\partial_x\partial_y
        - \frac{f'^2(r)}{4} e^{2i\theta}\Bigr] (i\sigma^y)
    \nonumber\\
    &\quad
    - \Delta_d f'(r)\,
    e^{-i\sigma^z\theta}\,
    \bigl[\cos\theta(\partial_x - i\partial_y)
        - \sin\theta(\partial_y - i\partial_x)\bigr].
    \label{eq:EffPair_id_final}
\end{align}

The first group of terms in each expression represents an effective spin-singlet pairing with odd angular momentum, while the final line in each expression is an induced spin-triplet pairing term with even angular momentum, whose amplitude scales linearly with \(f'(r)\). Therefore, the competition controlling topology is determined by the amplitude ratio:

\[
\boxed{\text{Phase transition occurs when}\quad
|\Delta_{\text{triplet}}| \sim |\Delta_{\text{singlet}}|.}
\]

In the semiclassical (Andreev) approximation, we can replace the derivatives with the the corresponding quasiparticle momentum, that is, $\partial_x \rightarrow i k_F \cos \varphi_{\mathbf{k}}, \partial_y \rightarrow i k_F \sin \varphi_{\mathbf{k}} $ to obtain

\begin{align}
\Delta_{\rm eff}^{(d+is)}(\mathbf{r},\mathbf{k})
&=
\Bigl[
\Delta_d k_F^2 \cos(2\varphi_{\mathbf{k}})
+ i\Delta_s
-\frac{\Delta_d f'(r)^2}{4}\cos(2\theta)
\Bigr]\, i\sigma^y
\nonumber\\
&\quad
- i k_F \Delta_d f'(r)\,
e^{-i\sigma^z\theta}\,
\cos(\theta+\varphi_{\mathbf{k}}),
\label{eq:Deltaeff_semiclass_dis}
\\[6pt]
\Delta_{\rm eff}^{(d+id)}(\mathbf{r},\mathbf{k})
&=
\Delta_d k_F^2 e^{2i\varphi_{\mathbf{k}}}\, i\sigma^y
-\Delta_d\,\frac{f'(r)^2}{4}\,e^{2i\theta}\, i\sigma^y
\nonumber\\
&\quad
- i k_F \Delta_d f'(r)\,
e^{-i\sigma^z\theta}\,
\Bigl[\cos\theta\,e^{-i\varphi_{\mathbf{k}}}
+i\sin\theta\,e^{i\varphi_{\mathbf{k}}}\Bigr].
\label{eq:Deltaeff_semiclass_did}
\end{align}

 
\subsection{Projection into the lower spin-orbit coupled band}
\label{app:projection_lower_band}

In this subsection we project the effective pairing matrix
$\Delta_{\mathrm{eff}}(\mathbf{r},\mathbf{k})$ onto the lower spin--orbit--coupled band
of the rotated normal-state Hamiltonian. Throughout we distinguish the real-space
polar angle $\theta$ (of $\mathbf{r}$) from the momentum-space polar angle
$\varphi_{\mathbf{k}}$ (of $\mathbf{k}$). The semiclassical expressions for
$\Delta_{\mathrm{eff}}^{(d+is)}$ and $\Delta_{\mathrm{eff}}^{(d+id)}$ used below are
given in Eqs.~\eqref{eq:Deltaeff_semiclass_dis} and
\eqref{eq:Deltaeff_semiclass_did}.

\paragraph*{Rotated normal-state Hamiltonian and eigen-spinors.}
In the bulk, retaining the dominant Rashba-like term generated by the skyrmion texture,
the rotated normal-state Hamiltonian takes the form
\begin{equation}
    h_0(\mathbf{k})
    = \xi_{\mathbf{k}} + \beta\,(k_y\sigma^x - k_x\sigma^y) + \alpha\,\sigma^z
    \equiv \xi_{\mathbf{k}} + \mathbf{h}_{\mathbf{k}}\cdot\boldsymbol{\sigma},
    \label{eq:h0_band_clean}
\end{equation}
with $\xi_{\mathbf{k}}=\frac{k^2}{2m}-\mu$ and
\begin{equation}
    \mathbf{h}_{\mathbf{k}}=(\beta k_y,\,-\beta k_x,\,\alpha),\qquad
    h_{\mathbf{k}}=|\mathbf{h}_{\mathbf{k}}|=\sqrt{\alpha^2+\beta^2 k^2}.
\end{equation}
Define angles $(\vartheta_{\mathbf{k}},\varphi_{\mathbf{k}})$ via
\begin{equation}
    \cos\vartheta_{\mathbf{k}}=\frac{\alpha}{h_{\mathbf{k}}},\qquad
    \sin\vartheta_{\mathbf{k}}=\frac{\beta k}{h_{\mathbf{k}}},\qquad
    e^{i\varphi_{\mathbf{k}}}=\frac{k_x+i k_y}{k}.
\end{equation}
A convenient choice of normalized eigen-spinors of $\mathbf{h}_{\mathbf{k}}\cdot\boldsymbol{\sigma}$
is
\begin{align}
    \ket{u_+(\mathbf{k})}
    &= \begin{pmatrix}
        \cos(\vartheta_{\mathbf{k}}/2) \\
        i\,e^{-i\varphi_{\mathbf{k}}}\sin(\vartheta_{\mathbf{k}}/2)
    \end{pmatrix},
    \\
    \ket{u_-(\mathbf{k})}
    &= \begin{pmatrix}
        i\,e^{i\varphi_{\mathbf{k}}}\sin(\vartheta_{\mathbf{k}}/2) \\
        \cos(\vartheta_{\mathbf{k}}/2)
    \end{pmatrix},
\end{align}
corresponding to eigenvalues $\xi_{\mathbf{k}}\pm h_{\mathbf{k}}$, respectively.

Let $\psi(\mathbf{k})=(\psi_\uparrow(\mathbf{k}),\psi_\downarrow(\mathbf{k}))^{\mathsf T}$ and
$\psi_{\rm band}(\mathbf{k})=(\psi_+(\mathbf{k}),\psi_-(\mathbf{k}))^{\mathsf T}$.
Then collecting the eigen-spinors into the unitary matrix
$U_{\mathbf{k}}\equiv (u_+(\mathbf{k}),u_-(\mathbf{k}))$, 
\begin{equation}
\psi(\mathbf{k}) = U_{\mathbf{k}}\,\psi_{\rm band}(\mathbf{k}),
\qquad
\psi_{\rm band}(\mathbf{k}) = U_{\mathbf{k}}^\dagger\,\psi(\mathbf{k}).
\label{eq:spin_band_transform_clean}
\end{equation}

The pairing term can be incorporated as
\begin{equation}
H_\Delta=\frac{1}{2}\sum_{\mathbf{k}}
\psi^\dagger(\mathbf{k})\,\Delta_{\mathrm{eff}}(\mathbf{r},\mathbf{k})\,\psi^\dagger(-\mathbf{k})
+\mathrm{H.c.},
\end{equation}
where $\Delta_{\mathrm{eff}}$ is a $2\times 2$ matrix in spin space.
Projecting onto the lower band yields
\begin{equation}
H_{\Delta,--}=\frac{1}{2}\sum_{\mathbf{k}}
\psi_-^\dagger(\mathbf{k})\,\Delta_{--}(\mathbf{r},\mathbf{k})\,\psi_-^\dagger(-\mathbf{k})
+\mathrm{H.c.},
\end{equation}
with the intraband pairing amplitude
\begin{equation}
\Delta_{--}(\mathbf{r},\mathbf{k})
=
u_-^\dagger(\mathbf{k})\,
\Delta_{\mathrm{eff}}(\mathbf{r},\mathbf{k})\,
u_-^*(-\mathbf{k}).
\label{eq:Dmm_projection_def_clean}
\end{equation}

\paragraph*{General closed form for $\Delta_{--}$.}
It is useful to decompose the pairing matrix as
\begin{equation}
\Delta_{\mathrm{eff}}(\mathbf{r},\mathbf{k})
=
A(\mathbf{r},\mathbf{k})\,i\sigma^y
+ B(\mathbf{r},\mathbf{k})\,\mathbbm{1}
+ C(\mathbf{r},\mathbf{k})\,\sigma^z.
\label{eq:Delta_decomp_ABC_clean}
\end{equation}
Evaluating Eq.~\eqref{eq:Dmm_projection_def_clean} with the eigen-spinor $u_-(\mathbf{k})$ above gives
\begin{align}
\Delta_{--}(\mathbf{r},\mathbf{k})
&=
-i\,A(\mathbf{r},\mathbf{k})\,\sin\vartheta_{\mathbf{k}}\,e^{-i\varphi_{\mathbf{k}}}
+ B(\mathbf{r},\mathbf{k}) - \cos \vartheta_{\mathbf{k}} C(\mathbf{r},\mathbf{k})
\label{eq:Dmm_full_no_largealpha}
\end{align}
with $\cos\vartheta_{\mathbf{k}}=\alpha/h_{\mathbf{k}}$ and $\sin\vartheta_{\mathbf{k}}=\beta k/h_{\mathbf{k}}$.

\paragraph*{$A,B,C$ for the present models.}
From Eqs.~\eqref{eq:Deltaeff_semiclass_dis}--\eqref{eq:Deltaeff_semiclass_did} we read off:
\begin{itemize}
\item For $d+is$,
\begin{align}
A_{d+is} &=
\Delta_d k_F^2 \cos(2\varphi_{\mathbf{k}})
+ i\Delta_s
-\frac{\Delta_d f'(r)^2}{4}\cos(2\theta),
\nonumber \\
B_{d+is} &=- i k_F \Delta_d f'(r)\cos(\theta+\varphi_{\mathbf{k}}) \cos\theta,\nonumber \\
C_{d+is} &=-k_F \Delta_d f'(r)\cos(\theta+\varphi_{\mathbf{k}}) \sin\theta.
\end{align}

\item For $d+id$,
\begin{align}
A_{d+id} &=
\Delta_d k_F^2 e^{2i\varphi_{\mathbf{k}}}
-\Delta_d\frac{f'(r)^2}{4}e^{2i\theta}, \nonumber \\
B_{d+id} &=- i k_F \Delta_d f'(r)\Bigl[\cos\theta\,e^{-i\varphi_{\mathbf{k}}}
+i\sin\theta\,e^{i\varphi_{\mathbf{k}}}\Bigr] \cos\theta,\nonumber \\
C_{d+id }&=- k_F \Delta_d f'(r)\Bigl[\cos\theta\,e^{-i\varphi_{\mathbf{k}}}
+i\sin\theta\,e^{i\varphi_{\mathbf{k}}}\Bigr]\sin\theta.
\end{align}
\end{itemize}
Substituting these coefficients into Eq.~\eqref{eq:Dmm_full_no_largealpha} yields the explicit
$\Delta_{--}(\mathbf{r},\mathbf{k})$ for each pairing symmetry.

\paragraph*{Large-$\alpha$ limit (optional simplification).}
For $\alpha\gg \beta k_F$ one has $\cos\vartheta_{\mathbf{k}}\simeq 1-\tfrac{1}{2}(\beta k_F/\alpha)^2$ and
$\sin\vartheta_{\mathbf{k}}\simeq \beta k_F/\alpha$. To $\mathcal{O} (\beta k_F / \alpha)$,  Eq.~\eqref{eq:Dmm_full_no_largealpha} simplifies as
\begin{equation}
\Delta_{--}(\mathbf{r},\mathbf{k})
\simeq
- i\,A(\mathbf{r},\mathbf{k})\,\frac{\beta k_F}{\alpha}\,e^{-i\varphi_{\mathbf{k}}}
+\bigl[B(\mathbf{r},\mathbf{k})-C(\mathbf{r},\mathbf{k})\bigr],
\end{equation}
which is the form used for simple analytical estimates of when $\Delta_{--}$ develops zeros on the
Fermi surface. 
\section{Low energy non-Majorana zero modes}\label{app:edge_modes}

In this appendix, we explore when low energy non-MZMs appear in our system, and explain why we use left and right Majorana overlap as our way of to diagnose the topological phase transition instead of the two other mentioned methods. 

\begin{figure}[t]
\begin{centering}
\includegraphics[width=3.6in]{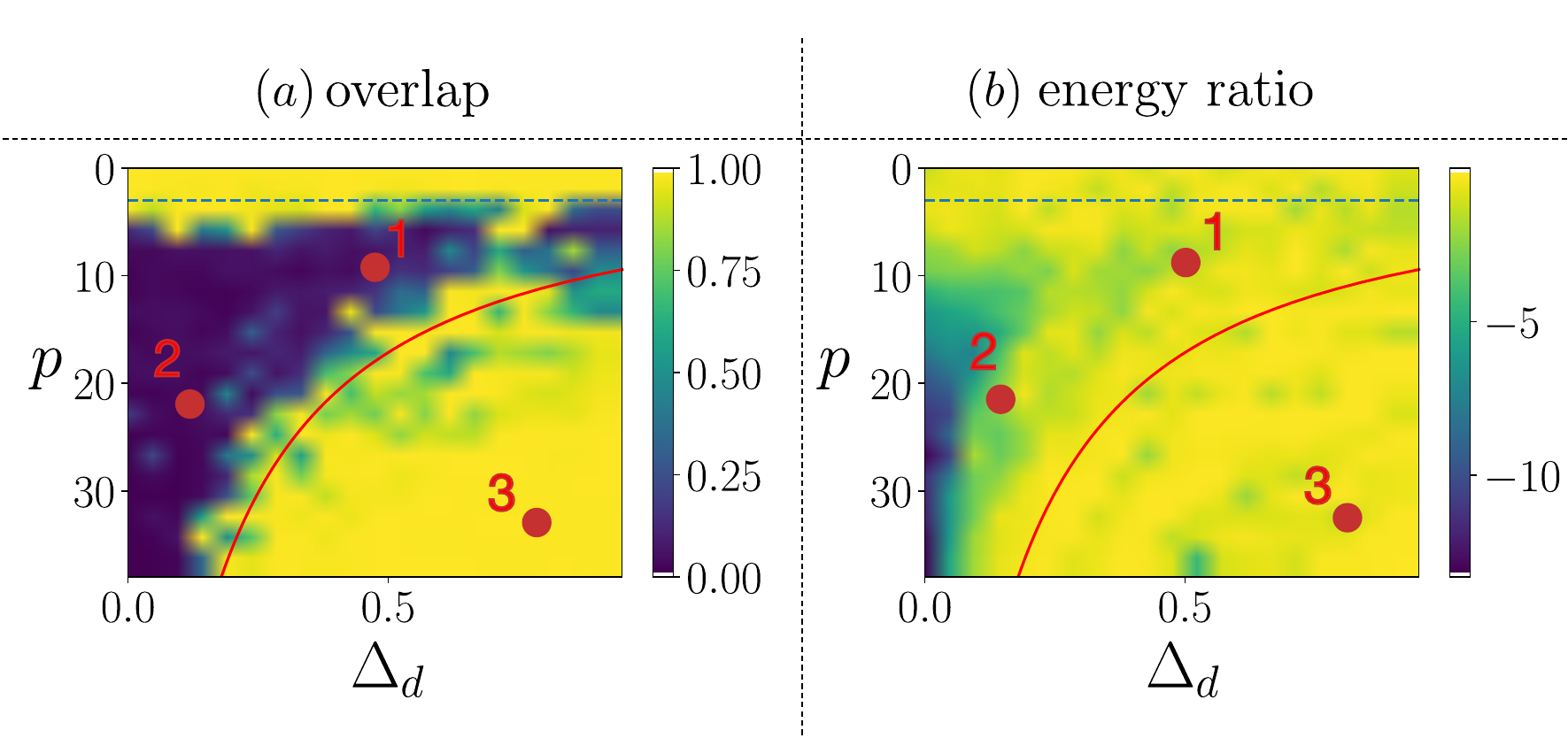}
\caption{Topological phases predictions for the d+is skyrmion model. (a) numerical overlap of left and right Majorana for the states closest to zero energy.(b) are the d+is phase diagrams using the ratio of energy closest to zero energy and the next energy level as an order parameter. The base of the logarithm for the energy plot is $10$. Points $1$ and $2$ on the phase diagram can be considered topological or non-topological, respectively, looking at the energy, but are clearly both in the topological regime when considering the overlap method. Point $3$ on the phase diagram is correctly judged to be in the non-topological phase via both methods.}
\label{fig:phase_diag_append}
\end{centering}
\end{figure}

Consider the phase diagrams for the $d+is$ case where we use energy and the overlap as order parameters. We pick a point on the phase diagram where the overlap method predicts that there are MZMs, but the energy method does not. Lastly we also pick a point which is predicted to be non-topological in both methods.

\begin{figure}
\begin{centering}
\includegraphics[width=2.5in]{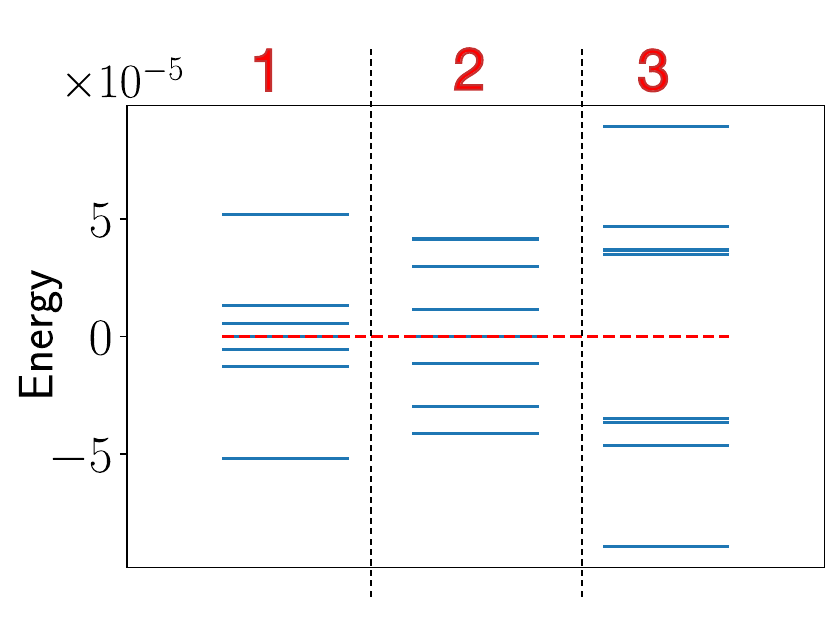}
\caption{Plots of first eight energies closest to zero energy. 1, 2 and 3 refer to the chosen points in Fig.~\ref{fig:phase_diag_append} on the phase diagram. }
\label{fig:energy_plot_append}
\end{centering}
\end{figure}

Plotting the energies in figure 5 for the three points, it is clear that visually speaking, 1. and 2. have a set of states very close to zero energy, while 3. does not have any states close to zero energy. 

However, referring to the phase diagram, we see that 1. should not be topological according to figure 4 b). However, this is simply because the MZMs worsen as we move away from 2.. At 2., the MZMs have an energy $3.8 \times 10^{-15}$ and for 1., the MZMs have an energy $2.8 \times 10^{-8}$. The energies away from zero energy are of the order of $10^-5$. Hence, when comparing the MZMs energies and higher energy states, for 1., the ratio of energies is a lot closer to what is expected from the non-topological phase. Hence, energy is less reliable as an order parameter as compared to the overlap between left and right Majoranas that we used.

\bibliographystyle{apsrev4-2}
\bibliography{Skyrmion}
\end{document}